\documentclass[pre,showpacs,showkeys,%
amsmath,amssymb,twoside,twocolumn,floatfix]%
{revtex4}

\usepackage[dvips]{graphicx}
\usepackage{exscale}
\usepackage{rotating}
\usepackage{bm}
\usepackage{time}
\usepackage{fancyhdr}

\newcommand{\fig}[1]{\ref{fig:#1}}
\newcommand{\Fig}[1]{Fig.~\ref{fig:#1}}
\newcommand{\Figs}[1]{Figs.~\ref{fig:#1}}
\newcommand{\eq}[1]{(\ref{eq:#1})}
\newcommand{\Eq}[1]{Eq.~(\ref{eq:#1})}
\newcommand{\Eqs}[1]{Eqs.~(\ref{eq:#1})}
\newcommand{\sect}[1]{Sec.~\ref{sec:#1}}

\newcommand{\App}[1]{App.~\ref{app:#1}}

\def\beq{\begin{equation}}
\def\eeq{\end{equation}}
\def\beqa{\begin{eqnarray}}
\def\eeqa{\end{eqnarray}}

\newcommand{\ie}{\emph{i.\,e., }}
\newcommand{\cf}{\emph{cf.\ }}
\newcommand{\rmd}{{\mathrm{d}}}

\def\6{\partial}

\newcommand{\free}{{\mathcal{F}}}
\def\freeK{K}

\def\vphi{\varphi}
\def\comA{\textsf{A}}
\def\comB{\textsf{B}}

\renewcommand{\vec}[1]{\mathbf{#1}}

\begin{document}


\title{Phase Separation under Ultra-Slow Cooling:
Onset of Nucleation}

\date{\today --- \now}

\author{J.\ Vollmer}
\email{juergen.vollmer@ds.mpg.de} %
\homepage{http://www-dcf.ds.mpg.de/Jvollmer} %
\affiliation{Fachbereich Physik, Philipps Universit\"at, 35032 Marburg, Germany
\\
Max Planck Institute for Dynamics and Self-Organization, Bunsenstr.~10, D-37073 G\"ottingen, Germany} %
%

\begin{abstract}
  We discuss the interplay between a slow continuous drift of
  temperature, which induces continuous phase separation, and the
  non-linear diffusion term in the $\phi^4$-model for phase separation
  of a binary mixture. This leads to a bound for the stability of
  diffusive demixing. It is argued that these findings are not
  specific to the $\phi^4$ model, but that they always apply up to
  slight modifications of the bound.  In practice stable diffusive
  demixing can only be achieved when special precautions are taken in
  experiments on real mixtures.  Therefore, the recent observations on
  complex dynamical behavior in such systems should be considered as
  a new challenge for understanding \emph{generic} features of
  phase-separating systems.
\end{abstract}



\maketitle

\pagestyle{fancy}
\fancyhead[LE]{\slshape J. Vollmer}
\fancyhead[RE]{\tiny \today $ $ --- \now}
\fancyhead[RO]{\slshape {Phase Separation under Ultra-Slow Cooling:
Onset of Nucleation}}
\fancyhead[LO]{\tiny \today $ $ --- \now}


\section{Introduction}

The kinetics of phase separation
\cite{LifSly61,Siggia79,Gunton83,Bray94,Binder98,onukiBook} is a topic of
continuous experimental
\cite{Tanaka94,tanaka95PRL,vollmer97EPL,vollmer97JCP1,vollmer97JCP2,vollmer99} and
theoretical interest \cite{WagYeo98,kendon99,puri01}. Many of its
characteristic features do not depend on specific materials, 
but are universal in the sense that they can be understood based on 
relatively simple model systems. 
%
%
For binary mixtures an appropriate model is the $\phi^4$ model
\cite{Bray94,Binder98,chaikinBook}. In the present paper we
augment this model by terms taking into account a slow continuous
drift of the temperature, in order to explore how the composition of
coexisting phases keeps up with the driving due to the temperature
evolution.

\begin{figure}
\[ \includegraphics[width=0.37\textwidth]{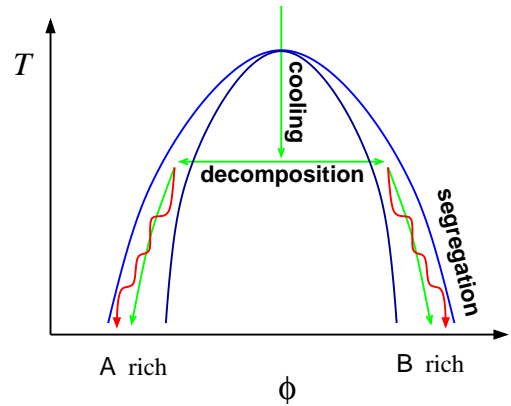} \]
\vspace*{-8mm}
\caption[phase diagram]{(Color online) 
  Schematic phase diagram for the demixing of a binary mixture. The
  binodal is given in blue and the spinodal line in dark blue.  A
  typical evolution of the composition $\phi$ of the mixture upon
  constant cooling is indicated by the green line. The red line shows
  an oscillating demixing, as it may be observed when the diffusive
  decomposition becomes unstable.
\label{fig:PhasDia}}
\end{figure}

Typically the demixing dynamics of binary fluids is idealized as a
three step process:
\\
A. \ A rapid change of the temperature (pressure or other experimental
control parameter) transfers the system from a single phase
equilibrium into the region, where phase separation is expected
[{\textsf{cooling}} in \Fig{PhasDia}]. This is supposed to be
sufficiently fast to neglect noticeable changes in the composition.
\\
B. \ The mixture decomposes into two coexisting macroscopic phases
[{\textsf{decomposition}} in \Fig{PhasDia}]. In applications this is
supposed to be fast on the time scales of changing the temperature;
in high precision experiments
\cite{wong81,chou81,joshua85,Tanaka94,tanaka95PRL,cumming90} a
temperature jump is applied and the temperature is subsequentially
fixed to carefully observe the dynamics in this regime.
\\
C. \ Upon further slow cooling there is an exchange of material such
that the composition of the coexisting phases follows the equilibrium
ones up to a small time lag [smooth green line denoted
{\textsf{segregation}} in \Fig{PhasDia}].

A dimensional analysis of the diffusive relaxation time
\cite{cates03,auernhammer05JCP,vollmer07PRL} shows, however, that for
experimentally relevant cooling rates the composition can not remain
close to the equilibrium values unless temperature is ramped
exceedingly slowly.
Indeed, recent experimental
\cite{sparks93,vollmer97EPL,vollmer97JCP1,vollmer99,vollmer97JCP2,HeimburgMirzaevKaatze2000,vollmer02PCCP,turchanin04CPL,turchanin04JCP,auernhammer05JCP}
and numerical studies \cite{wagner03Proceedings,cates03} indicate that
rather than slowly following the drift of the equilibrium composition,
often there is very complex dynamics observed in regime C. Among
others secondary nucleation in large domains (\emph{double phase
  separation}) \cite{tanaka95PRL}, oscillating bursts of massive phase
separation alternating with quiescent periods
\cite{vollmer97EPL,vollmer97JCP1,vollmer99,vollmer97JCP2,HeimburgMirzaevKaatze2000,vollmer02PCCP,turchanin04CPL,turchanin04JCP}
and stationary convection patterns \cite{cates03} have been reported.
Except for the stationary convection, all these scenarios give rise to
an explicitly time dependence of the composition like the one
indicated by the red line in \Fig{PhasDia}.

Despite the impressive experimental findings, the theoretical
understanding of such complex dynamical response to slow cooling is
still at its infancy. To make a point, however, on the prevalence of
the various complex patterns of phase separation the present paper
explores the critical cooling rate beyond which stable diffusive
demixing can no longer be expected.

The article is organized as follows.
Section \sect{EOM} revisits the relevant nonlinear diffusion equation
describing the evolution of concentration in a slowly cooled binary
mixture. 
In \sect{Stable} we consider the limit of a very narrow interface at
the meniscus in order to analytically derive a phase portrait
which gives us access to discuss stable and unstable solutions of the
diffusion equation, and their bifurcations. A generalization of this
discussion to the case where the phases are not symmetric is given in
\App{asymm}. 
The subsequent \sect{galerkin} introduces a mode expansion in order to
gain insight in the time evolution of the profiles after bifurcations,
and the role of higher order derivatives needed to properly describe
interfaces of finite thickness. (Technical issues of its derivation
have been delegated to \App{modeExpansion}.)
The main results are summarized in \sect{discussion}.

\begin{figure*}
\[ 
   \includegraphics{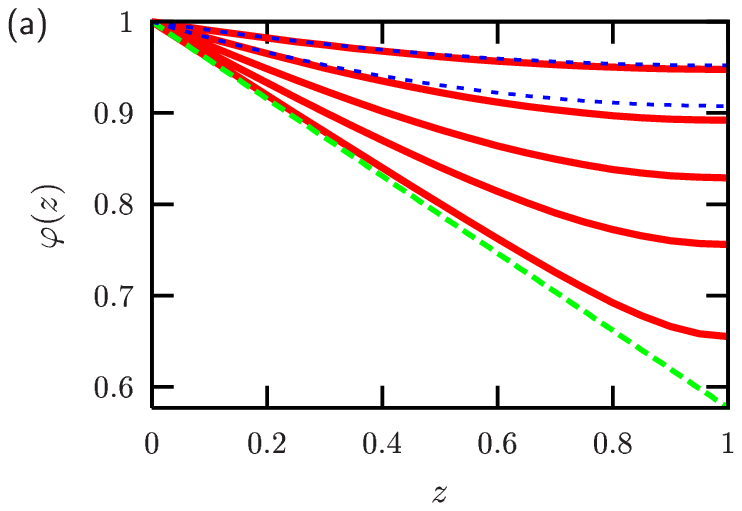} 
   \qquad\quad\quad
   \includegraphics{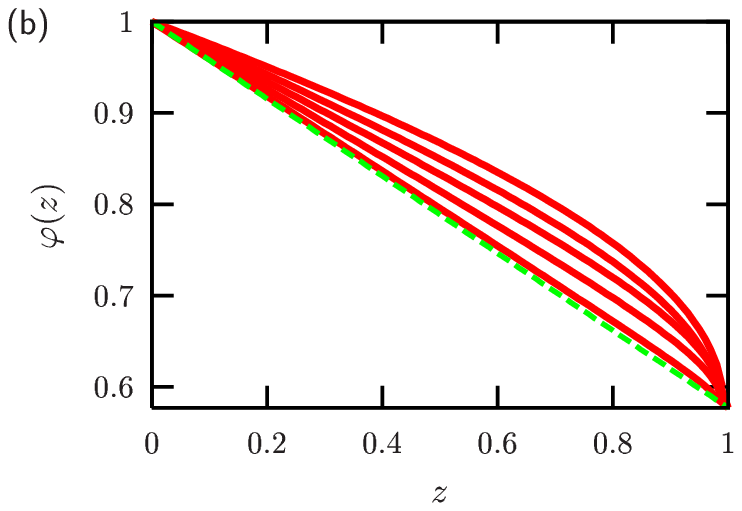} 
\]
\vspace*{-6mm}
\caption[profiles]{(Color online) 
  Stationary density profiles for $M=0$ and
  $N = 0.1, 0.2, 0.3, 0.4, 0.50$ (solid red lines from
  top to bottom), and the critical parameter 
  $N_c = (\sqrt{3}-1)^2 \simeq 0.5359$ (dashed green curve): 
  (a) stable solutions with $\vphi'(z=1)=0$; 
  (b) unstable solutions, which all end at $\vphi_s = 3^{-1/2} \simeq 0.58$.   
  The dotted blue lines in part (a) show the approximation
  \eq{profApprox} for $N = 0.1$ and $N = 0.2$, respectively. 
\\
\label{fig:profiles}}
\end{figure*}
\section{Nonlinear Diffusion Equation}
\label{sec:EOM}

We describe the temporal evolution of the binary mixture by the
free energy functional \cite{chaikinBook}
\begin{equation}
   \free[\phi(\vec{x};t), t] = \int \rmd \vec{x}
        \left[ F_0[\phi(\vec{x};t)] 
             + \frac{\freeK}{2} \: \left(\nabla \phi(\vec{x};t) \right)^2 \right]
\label{eq:FreeDens}
\end{equation}
where $\phi(x)$ characterizes the composition at position $x$, and the
integration over $\vec{x}$ is performed over the volume of the system. The
second term in square brackets represents an energy penalty for steep
changes of composition (macroscopically it gives rise to surface
tension), and
\begin{eqnarray}
    F_0(\phi)
& \equiv &
    \frac{B}{4} \phi^4 + \frac{A}{2} \phi^2
\nonumber\\
& = &
    -A \phi_0^2 \left[
        \frac{1}{4} \left(\frac{\phi}{\phi_0}\right)^4
      - \frac{1}{2} \left(\frac{\phi}{\phi_0}\right)^2
    \right]
\label{eq:FreeEn}
\end{eqnarray}
is the free energy of the $\phi^4$-model, which describes the
equilibrium phase behavior of the mixtures. In the present study we
only deal with negative values of $A$, where there are two
coexisting phases with composition $\pm \phi_0$ and $\phi_0 \equiv
\sqrt{-A/B}$.  The temperature dependence of the equilibrium
composition $\phi_0(T)$ is due to the temperature dependence of the
parameters $A$ and $B$.

The free energy \eq{FreeEn} faithfully accounts for qualitative
features of the demixing of fluid mixtures, even though more
complicated expressions may arise for concrete systems
\cite{Bray94,chaikinBook}.
The explicit time dependence in \Eq{FreeDens} accounts for the fact
that we deal with a system subjected to a sustained change of
temperature.  Due to its temperature dependence the equilibrium
composition $\phi_0$ becomes explicitly time dependent in that case.

In the absence of center of mass flow the interdiffusion of the
components of the mixtures follows
\[
   \6_t \phi(\vec{x};t) = \alpha \nabla^2 \mu[\phi(\vec{x};t)]
\]
where $\alpha$ is the thermodynamic coefficient for mass
interdiffusion, and the chemical potential
$\mu[\phi]
 \equiv
\delta (F_0[\phi]+\freeK/2 \, [\nabla\phi]^2) / \delta\phi$
is the functional derivative of the free energy density \eq{FreeDens}
with respect to $\phi$.
Consequently, the normalized composition $\vphi\equiv\phi/\phi_0$
evolves according to the nonlinear diffusion equation
\begin{equation}
   \6_t \vphi
 = \nabla\left[ \frac{D}{2} \left(3\vphi^2-1\right) \: \nabla\vphi\right]
   - D \delta^2 \, \nabla^4 \vphi
   - \xi\, \vphi
\label{eq:NLdiffuEq}
\end{equation}
where 
\begin{equation}
\xi\equiv \frac{ \6_t\phi_0 }{ \phi_0 }
\label{eq:xi}
\end{equation}
accounts for the change of the equilibrium composition due to the
changing temperature, $\delta\equiv (\freeK/|A|)^{1/2}$ is the width
of a planar macroscopic interface between the coexisting phases
\cite{chaikinBook}, and $D\equiv -\alpha A$ is the equilibrium
diffusion coefficient (\ie the one for a system with uniform
composition $\vphi\equiv \pm 1$).  Away from equilibrium the diffusion
becomes nonlinear with a diffusion coefficient $D\,(3\vphi^2-1)/2$
explicitly depending on composition.  In the \emph{spinodal} range
$-\vphi_s \equiv -1/\sqrt{3} < \vphi < 1/\sqrt{3} \equiv \vphi_s$
the diffusion coefficient even takes negative values, indicating the
instability of the mixture against arbitrarily small fluctuations.

In the following we focus on the simplest setting allowing us to
discuss the breakdown of diffusion --- generalizations are outlined in
\App{asymm}:
\\
(i) There are equal amounts of the components \comA\ and \comB\ of the
mixture such that in equilibrium there is a macroscopic \comA-rich
phase coexisting with a \comB-rich one.
\\
(ii) The mixture is kept in a container of constant cross section and
height $2\Lambda$, and it is subjected to a gravitational field. As a
consequence there is a macroscopic phase boundary at $z=0$. The
bottom and top walls of the container are at $z=\pm\Lambda$,
respectively.
\\
(iii) The flux through the bottom and top of the system vanishes.
\\
(iv) There is no flow in the horizontal directions such that
$\6_x\vphi = \6_y\vphi = 0$.
\\
(v) The temperature is changed in such a way that $\xi$ is constant
(\cf\cite{auernhammer05JCP}).
\\
(vi) The temperature and composition dependence of the equilibrium
diffusion coefficient $D$ may be neglected.

Spacial scales are conveniently measured in units of
$\Lambda$, and time in units of $\Lambda^2/D$. The
resulting dimensionless diffusion equation
\begin{equation}
\6_t \vphi
=
   \6_z\left[ \frac{3\vphi^2-1}{2} \: \6_z\vphi \right]
 - M^2 \: \6_z^4 \vphi
 - N \: \vphi
\label{eq:diffu}
\end{equation}
takes a universal form for \emph{all} the mixtures. It only involves the
dimensionless parameters
\begin{subequations}
\begin{eqnarray}
  N &\equiv& \frac{\xi \Lambda^2}{D} ,
\label{eq:defN}
\\
  M &\equiv& \frac{\delta}{\Lambda} .
\label{eq:defM}
\end{eqnarray}
\end{subequations}
The former characterized the strength of the driving by the ratio of
the rate of generation $\xi$ of supersaturation over the rate of decay
of supersaturation due to diffusion. The latter characterizes the
importance of the boundary layer at the meniscus between the
coexisting phases by the ratio of the interface width $\delta$ over
the system size $\Lambda$.

Equation \eq{diffu} ought to be solved subject to no-flux boundary
conditions (iii) at the top and bottom of the system, viz.
\begin{equation}
   0 = \left[\,
          \frac{3\vphi^2-1}{2} \: \6_z\vphi
        - M \: \6_z^3 \vphi \,
        \right]_{z=\pm 1} .
\label{eq:BC}
\end{equation}
To ensure mass overall conservation we furthermore demand
$\vphi(z)=-\vphi(z)$ \footnote{From the symmetry condition
  $\vphi(z)=-\vphi(z)$ we may infer $\vphi(0)=0$ and
  $-\vphi(-1)=\vphi(1)$, such that this requirement indeed provides us
  with four independent boundary conditions required to uniquely
  determine the stationary solution of the fourth order differential
  equation \eq{diffu}.}.

\section{Stationary diffusion profiles}
\label{sec:Stable}

Except in the critical region (\ie the region very close to the
maximum of the coexistence curve in \Fig{PhasDia}) the width $\delta$
of the interface is always much smaller than the lateral extension
$\Lambda$ of a macroscopic system.  Observing that a temperature ramp
rapidly moves the mixture out of this region, one may assume
$M \equiv \delta/\Lambda \ll 1$ in systems of macroscopic dimension $\Lambda$.
Consequently, the term with the fourth order derivative, which
accounts for the effects of surface tension, is only relevant in
situations where there are very rapid changes of composition.
This only is the case close to $z=0$ at the interface between the
coexisting phases, and for values of $\vphi$ close to the threshold of
spinodal decomposition (\cf below).
Typically it has only negligible influence on the supersaturation
profile $\vphi(z;t)$. To gain insight in its dependence on $N$
we therefore discuss the $\vphi(z)$ profiles in
the idealized case $M=0$, where the diffusion equation
\begin{equation}
\6_t \vphi
=
   \6_z\left[ \frac{3\vphi^2-1}{2} \: \6_z\vphi \right]
 - N \: \vphi
\label{eq:profile}
\end{equation}
will be subjected to the boundary conditions 
\begin{subequations}
\begin{eqnarray}
  z=0^{\pm} : &\quad& \qquad\qquad\qquad
  \vphi(z=0) = \pm 1 ,
\label{eq:bc0}
\\
  z = \pm 1  : &\quad&
  \frac{3\vphi^2-1}{2} \: \6_z\vphi(z=\pm 1;t) = 0 .
\label{eq:bc1}
\end{eqnarray}
\end{subequations}
The former requires that the the composition takes on the equilibrium
value at the meniscus, and the latter that there is no flux through
the top and bottom interfaces of the system.
There are two possibilities to fulfill the latter boundary condition (\Fig{profiles}):
\\
(a) stable diffusion profiles are obtained for 
$\6_z\vphi(z=\pm 1) = 0$; 
\\
(b) when the composition at the top and/or bottom of the system
exactly amounts to the spinodal composition 
$\vphi(z=\pm 1) = \vphi_s = 3^{-1/2}$ 
the factor in front of the derivative vanishes, and one
obtains an unstable diffusion profile.
\\
These possibilities will be further explored in the remainder of this section. 
The question in how far the findings based on this idealized setting
still apply at a non-vanishing $M$ will be addressed in \sect{galerkin}.

\subsection{Stable diffusion profiles}
\label{sec:stableDiffProfile}

For very weak driving, $N \lll 1$, the composition $\vphi$ never
substantially differs from its equilibrium value, such that
$(3\vphi^2-1)/2 \simeq 1$. In other words the diffusion coefficient
hardly depends on $\vphi$, such that it may be approximated by its
equilibrium value. In this case the boundary condition \eq{BC} at the
top requires $\6_z\vphi=0$, and one easily checks that the stationary
solution of the diffusion equation amounts to
\begin{equation}
   \vphi(z) =
        \frac
        {\cosh\left[ \sqrt{N} \; (z-1)  \right]}
        {\cosh\left[ \sqrt{N}  \right]} .
\label{eq:profApprox}
\end{equation}
For $N=0.1$ and $N=0.2$ it is shown by a dotted blue line in
\Fig{profiles}(a). Indeed the approximation appears to be very good as
long as $N\lesssim 0.1$. 

For larger values of $N$ the $\vphi$-profile deviates too strongly from
unity to neglect the supersaturation dependence of the diffusion
coefficient. In that case
the solution of the diffusion equation \eq{diffu} can not be given in
a closed form.  Upon numerical integration one observes, however, that
close to $z=0$ it closer and closer approaches the linear solution
\begin{equation}
   \vphi(z) = 1 - \sqrt{ \frac{N}{3} } \; z .
\label{eq:LinSol}
\end{equation}
Unless $N = N_c\equiv (\sqrt{3} - 1)^2 \simeq 0.5359$ this linear
solution does not fulfill the no-flux boundary condition \eq{bc1} at
the surface $z=0$. Consequently, the solution eventually crosses over
into an approximately parabolic profile with a minimum at $z=1$ (solid
red lines in \Fig{profiles}(a)). 
In the limit $N \to N_c$ the stable solutions closer and closer
approaches the linear solution $1-\sqrt{N_c/3} \, x$, which for $N=N_c$ 
fulfills the boundary condition~\eq{BC} because the ratio preceding
$\6_z\vphi$ vanishes.


\begin{figure}
\[ 
\includegraphics{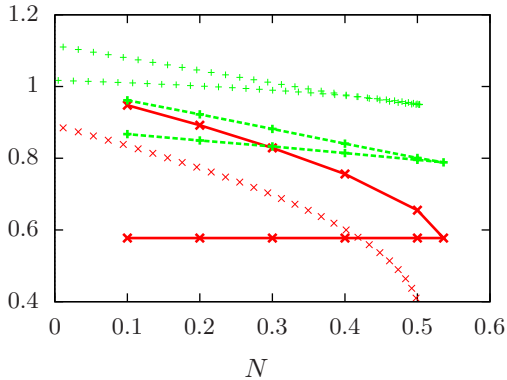}
\]
\vspace*{-6mm}
\caption[bifurcation]{(Color online) The values $\vphi(z=1/2)$ (thick
  green $+$ connected by dahsed lines) and $\vphi(z=1)$ (thick red $\times$
  connected by solid lines) of the stable (upper curve) and unstable (lower
  curve) solution as a function of the dimensionless ramp rate $N$.
  The respective approximations by the two mode model \Eq{2modeModel} are 
  indicated by corresponding small symbols.
  \label{fig:bifurcation}}
  The corresponding small symbols show 
\end{figure}

\subsection{Unstable stationary profiles and a saddle-node bifurcation}
\label{sec:Unstable}

%
From a physical point of view the linear solution for $N_c$
has no flux across the top surface since the nonlinear diffusion
coefficient $(3\vphi^2-1)D/2$ vanishes at the spinodal.
There are several observation to be made about this solution:
\\
(i) Since the composition close to the top of the system approaches
the value at the spinodal line, the linear profile is unstable against
nucleation of domains of the minority phase close to top end.
As a consequence the stable solutions of the diffusion equation
approach an \emph{unstable} solution, when $N \to N_c$.
\\
(ii) Besides being the limiting behavior of the stable solutions,
the linear profile can also be viewed as a limiting case of a family
of solutions that end on the spinodal point at the top of the system
(solid red lines in \Fig{profiles}(b)).  All these solutions are
unstable against nucleation of domains of the minority phase close to
top end.
\\
(iii) At the critical parameter value $N_c$ 
%
%
the family of stable solutions of the diffusion equation collides with
the unstable ones in a saddle node bifurcation. Subsequently, there
is no stationary diffusive solution with only two domains in the
vertical direction. In order to clearly make this point
\Fig{bifurcation} shows the values $\vphi(z=1/2)$ and $\vphi(z=1)$ for
the coexisting stable (upper curves) and unstable (lower curves)
solutions.

This result still holds for non-vanishing $M$, since $\6_z^4\vphi$
vanishes for a linear profile. 
Moreover, it also does not rely on the specific dependence $(3
\vphi^2-1)/2$ of the $\phi^4$ model. When the supersaturation
dependence of the diffusion is an even 2nd order polynomial $(a
\vphi^2-b)/(a-b)$ the stable and unstable profiles still approach a
linear profile which approaches the spinodal concentration $\vphi_s =
\sqrt{a/b}$ exactly at the critical point
\begin{equation}
  N_c = 2\; \frac{1 - \vphi_s}{1 + \vphi_s}
\label{eq:NcrExact}
\end{equation}
where the stable and the unstable profile merge and disappear in a
saddle node bifurcation. 
The representation of the stationary profiles via flow diagrams
provides us with more insight into the structure of the solutions, and
how they might change upon changes of the transport process.

\begin{figure}
\[ \includegraphics{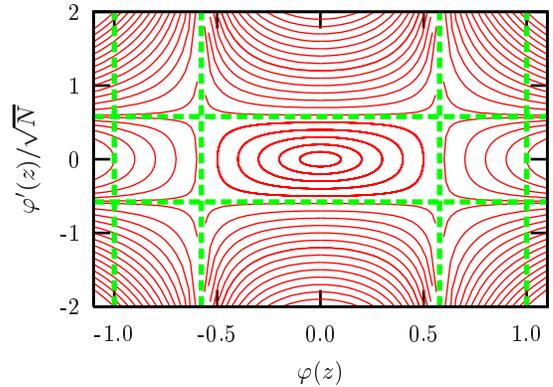} \]
\vspace*{-6mm}
\caption[phase flow]{(Color online)  Flow diagram of \Eq{profile} with
  $M=0$ and $N$ absorbed in the length scale (\cf text). The
  equilibrium ($\vphi=\pm 1$) and spinodal ($\vphi=\pm 1/\sqrt{3}$)
  compositions are marked by vertical dashed green lines. Note that
  the linear profiles \eq{LinSol} run horizontally on lines
  $\vphi'/\sqrt{N} = \pm 1/\sqrt{3}$, which are also indicated by
  dashed green lines. 
  \label{fig:phaseFlow}}
\end{figure}

\subsection{Flow diagram}
\label{sec:FlowDiagram}

The different types of stationary supersaturation profiles $\vphi(z)$
can conveniently be summarized by observing that in \Eq{profile} the
control parameter $N=\xi\Lambda^2/D$ can be absorbed into the length
scale, for instance by choosing 
$z\to \hat{z}\equiv z / \sqrt{N} = (D/\xi)^{1/2} \, z/\Lambda$ %
\footnote{ In this case the length scale is measured in terms of the
  natural diffusive scale $(D/\xi)^{1/2}$.}.
The equation \eq{profile} has no free parameter in that case, such
that each choice of initial conditions $(\vphi(1); \vphi'(1))$ leads
to a unique stationary solution of the diffusion equation. 
In other words, the solutions of \Eq{profile} at different $N$ \emph{only}
differ by a rescaling of the length scale. Consequently, all solutions
can be represented as a flow diagram in the $(\vphi, \vphi'/\sqrt{N})$
plane (\Fig{phaseFlow}). 

The flow is symmetric with respect to reflections at both coordinate
axes. 
The symmetry with respect to reflections at the horizontal
axis reflects the free choice of the orientation of the $z$ axis. The
inversion symmetry of the plot is a consequence of the symmetry
between the phases in the $\phi^4$ model. In \Eq{profile} these
symmetries amount to invariance under $z\to -z$ and $\vphi\to -\vphi$.
%
The symmetries are a consequence of the assumption that there are
equal amounts of \textsf{A} and \textsf{B} in the mixture, and that
the free energy is symmetric with respect to interchanging \textsf{A}
and \textsf{B}.
\App{asymm} demonstrates that our findings do not substantially change
in the non-symmetric case. There are only changes of unstable
trajectories residing entirely in the spinodal regime where $|\vphi|
< \vphi_s$. Up to this change the flow is structurally stable (\cf
\cite{Arnold2008}) upon variation of the functional form of the source
term (as long as it remains continuous and strictly monotonous) and
the dependence of the diffusion on the supersaturation (as long as one
replaces $(3\vphi-1)/2$ by a continuous convex function which takes
the value unity at the equilibrium compositions $\vphi=\pm 1$ and
negative values somewhere in between).

In order to identify the trajectories in \Fig{phaseFlow} with the
profiles shown in \Fig{profiles}, we observe that all solutions shown
in \Fig{profiles} have in common that $z_0=0$ and $\vphi(0)=1$. They
differ in the choice of the initial slope $\vphi'(0)/\sqrt{N}$.  There is a
stable solution with slope $\vphi'(1)$ slightly smaller than
$\sqrt{N/3}$, and an unstable solution where the slope is initially
somewhat larger. For stronger driving the initial slopes of the
solutions ever closer approach $\sqrt{N/3}$, and the solutions remain
linear for longer times. 

In the phase flow the profiles all correspond to trajectories which
start on the line $\vphi=1$ (rightmost dashed green line) and move inward
with a negative initial slope. If $0\geq \vphi'(0)/\sqrt{N} > -1/\sqrt{3}$ one
obtains a stable profile, which ends on the line $\vphi'=0$. For
$\vphi'(0) < -1/\sqrt{3}$ one obtains an unstable profile, which
terminates with infinite slope at the vertical green line
$\vphi=1/\sqrt{3}$, which indicates the spinodal $\vphi_s$.
For each choice of $N$ there is \emph{exactly} one
stable trajectory and one unstable trajectory, which fulfill these
boundary conditions for $z = 1$.
In addition there are symmetry-related solutions in the four other
quadrants of the diagram.

In addition to the supersaturation profiles, which start at $\vphi=1$
there are additional stationary states where the supersaturation
always lies within the spinodal regime. 
More complicated profiles can be build by concatenation of pieces of
profiles, provided that $(3\vphi^2-1)\, \vphi'(z)$ remains continuous in
order to guarantee a continuous diffusion current. At least in
principle, sharp changes of $\vphi$ are admissible because they are
the hallmark of phase boundaries.
However, even for negligibly small $M$ the forth order derivative term
comes into play here, because it suppresses steep gradients and jumps
in $\vphi$. Only the presence of this term makes it possible to
discuss stable and metastable diffusion profiles connecting \comA\
rich and \comB\ rich states.

\section{Mode expansion}
\label{sec:galerkin}

The relaxation towards 
the stable profiles 
is a dynamic process.
In order to also
gain insight into the structure of the stationary profiles and their
bifurcations, we resort to a mode expansion
\begin{equation}
  \vphi(z;t) = \frac{2}{\pi} \; {\sum_j}' \vphi_j (t) \; \sin \frac{\pi j z}{2}
\label{eq:expansion}
\end{equation}
where the prime at the sum indicates that the sum is to be taken over
all odd integers $j$. Moreover, we require $\vphi_{-j} = -\vphi_j$
such that this ansatz automatically fulfills the antisymmetry of the
composition profile (which we had required to ensure mass conservation)
and the boundary condition \eq{BC} of the diffusion equation
\eq{profile}.
In other words it shows an inversion symmetry at the phase boundary
$z=0$, and vanishing slope at the top and bottom $z=\pm 1$.

Truncating the expansion after the first few leading order modes will
provide us with a finite-dimensional dynamical system mimicking the
temporal dynamics of the system.

\subsection{Equations of motion}

Inserting \Eq{expansion} into \Eq{diffu} and using trigonometric
relations in order to express products of sines in sines
(\cf\App{modeExpansion}) leads to
\begin{eqnarray}
  \dot\vphi_j
  &=&
  -
  \left( N  -  \frac{\pi^2 j^2}{8} +  \frac{\pi^4 M^2 j^4}{16} \right)  \vphi_j
\nonumber\\&&
  +
  3 \, \sum_{kl} 
  \left(kl + \frac{k^2}{2} \right) \; 
  \vphi_k \vphi_l \vphi_{j-k-l}
\label{eq:dot_vphi}
\end{eqnarray}

\begin{figure*}
\[ \includegraphics{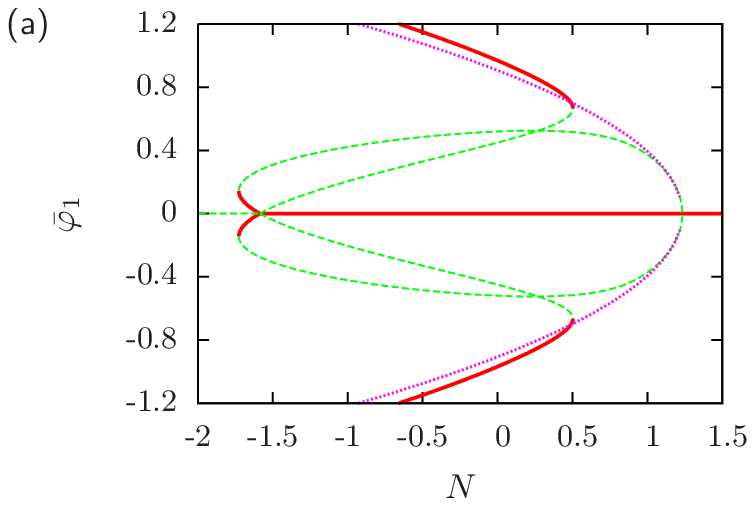} 
\qquad\quad\quad  
   \includegraphics{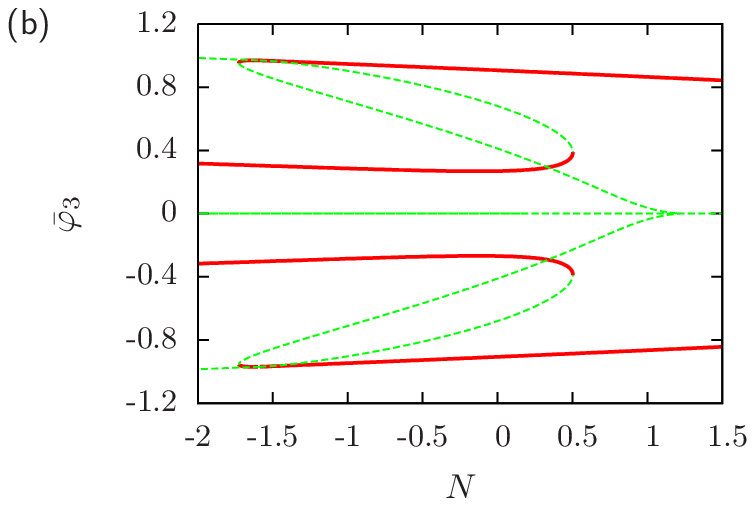} 
\]
\vspace*{-6mm}
\caption[galerkin]{(Color online) 
  The panels show the amplitudes (a) $\bar\vphi_1$ and (b)
  $\bar\vphi_3$ of the steady-state solutions of the two mode
  approximations to the concentration profile \Eq{2modeModel}. 
  Stable solutions are indicated by a solid red line, and the unstable
  ones by dashed green line. 
  For comparison the dotted purple line in panel (a) also shows the single
  mode approximation \Eq{barPhi}.
  \label{fig:amplitudes}}
\end{figure*}

\subsubsection{Single Mode}
Taking into account only the modes $j=\pm 1$ leads to 
\begin{equation}
  \dot\vphi_1
  =
  -
  \left( N  -  \frac{\pi^2}{8}  +  \frac{\pi^4 M^2}{16} \right)  \vphi_1
  -
  \frac{3}{2} \vphi_1^3
\label{eq:1modeModel}
\end{equation}
It admits the stationary solutions $\vphi=0$ and the nontrivial
solution 
\begin{equation}
\bar\vphi
= \frac{\pi}{2} \, \left( 
    \frac{1}{3} 
  - \frac{\pi^2 M^2}{6} 
  - \frac{8 N}{3 \pi^2} 
 \right)^{1/2}
\label{eq:barPhi}
\end{equation}
which ceases to exist at the critical point
\begin{equation}
N_c^{(1)} = \frac{\pi^2}{8} - \frac{\pi^4 M^2}{16}
\label{eq:Ncr1}
\end{equation}
Already in this simplest conceivable framework it is not possible to have
a stable diffusion profile when the parameter $N$ characterizing the
strength of the temperature drift becomes too large. 
However, in this setting the amplitude $\bar\vphi$ of the
supersaturation profile vanishes upon approaching $N_c$ rather than
that the stable profile vanishes in a saddle node bifurcation as
suggested by the findings summarized in \Fig{bifurcation}.

\subsubsection{Two Modes}
Taking into account the two leading order modes $j=\pm 1, \pm 3$ leads to 
\begin{widetext}
\begin{subequations}
\begin{eqnarray}
  \dot\vphi_1
  &=&
  -
  \left( N - \;\; \frac{\pi^2}{8} + \;\;\; \frac{\pi^4 M^2}{16} \right)  \vphi_1
  - \frac{3}{2} \vphi_1^3
  + \frac{3}{2} \vphi_1^2 \vphi_3
  - 3\, \vphi_1 \vphi_3^2
\\
  \dot\vphi_3
  &=&
  -
  \left( N
    - \frac{9\pi^2}{8} 
    + \frac{81\pi^4 M^2}{16} 
  \right)  \vphi_3
  + \frac{9 }{2} \vphi_1^3
  - 27\, \vphi_1^2 \vphi_3
  - \frac{27}{2} \vphi_3^3
\end{eqnarray}
\label{eq:2modeModel}
\end{subequations}
\end{widetext}
For the case of narrow interfaces $M=0$ its steady-state solutions are
plotted in \Fig{amplitudes}:
\\
There still is a solution whose amplitude continuously disappears upon
approaching the critical parameter value $N_c^{(1)} = \pi/8 \simeq
1.2337$ found in \Eq{Ncr1}. Close to the bifurcations its amplitudes
are very close to the one of the single mode approximation (purple
dotted line). However, this solution is now correctly identified as an
unstable solution where the concentration lies within the spinodal
region all the time. It corresponds to one of the closed orbits in the
center region of \Fig{phaseFlow}.
\\
Moreover, the two mode model admits two qualitatively new types of
solutions.
For all $N>0$ there is a stable solution $\bar\vphi_1=0$ and 
$\bar\vphi_3 = \frac{\pi}{2} \, \left( 
    \frac{1}{3} 
  - \frac{\pi^2 M^2}{6} 
  - \frac{8}{3 \pi^2} \; \frac{N}{9}
 \right)^{1/2}
$
which amounts to the single mode approximation of a system of height
$\Lambda/3$ rather than $\Lambda$.
\\
In addition, for small $N$ there are two (symmetry-related) pairs of
non-trivial solutions, which disappear in saddle-node bifurcations at
$N_c^{(2)} \simeq 0.5031$, which is remarkably close to the exact
value $N_c = (\sqrt{3}-1)^2 \simeq 0.5359$ obtained in
\Eq{NcrExact} (\cf \Fig{profiles}). 
Beyond this value of $N$ diffusion is too slow to remove the
supersaturation from a domain of size $\Lambda$, and the system
approaches the stable fixed point with vanishing amplitude
$\bar\vphi_1$. 

The saddle node bifurcation at $N_c$ and the subsequent decay towards
a solution with domains of size $\Lambda/3$ has a simple physical
interpretation. 
It amounts to nucleation of new domains close to the top and bottom of
the system where supersaturation is largest, such that the domain size
is reduced to one third of its previous value and diffusive relaxation
can again keep track with the driving of the system. 
When relaxing the constraint of translational invariance in horizontal
direction, nucleation would of course give rise to a cloud of droplets
rather than a full two-dimensional layer of material. A discussion of
the related instabilities and feedback-mechanisms lie outside the
scope of the present work. First steps to their understanding have
been suggested in \cite{vollmer07PRL}, and a more detailed analysis
will be given in forthcoming work.
At this point we conclude that the two-mode model faithfully describes
the breakdown of stable diffusion by a saddle-node bifurcation. Even
though the values of the supersaturation are reproduced only in a
crude approximation (dashed lines in \Fig{bifurcation}) its prediction
of the critical point is off by less than $7$\%. 
Adding more modes in first place gives a more realistic description of
the stable solutions remaining for $N>N_c$ and in the completely
analogous saddle node bifurcations leading to breakdown of diffusion
in these structures, when $N$ becomes still larger.

\section{Discussion}
\label{sec:discussion}

In this paper we have studied the diffusive transport in binary
mixtures in response to ultra-slow cooling. Starting from the generic
$\phi^4$ model of binary fluid mixtures we showed  
\begin{itemize}

\item  that stable diffusion is only possible up to a critical parameter
  value \Eq{NcrExact} where the stable diffusion profiles merges with
  an unstable profile, and disappears in a saddle-node bifurcation.

\item  This prediction of the critical value applies whenever the
  concentration dependence of diffusivity is well-approximated by an
  even 2nd order polynomial. 

\item  In the limit of sharp interfaces the steady state profiles may
  be represented in the form of flow diagrams \Figs{phaseFlow} and
  \fig{asymmPhaseFlow}. Since the flow is topologically stable, the
  findings are not restricted to the specific model but are robust to
  changes in the phase boundaries and concentration dependence of the
  diffusivity. 

\item  An expansion of the model in a Fourier series shows that the
  picture also applies when the forth order derivatives are included
  in the diffusion equation \eq{diffu} that are needed to account for
  the finite width of the interface between coexisting phases. 

\item When approaching the critical parameter value of the saddle-node
  bifurcation the concentration profile becomes unstable with respect
  to nucleation of droplets, which will break the horizontal
  translation invariance assumed in the present study. 

\end{itemize}
The two-mode approximation of the concentration profiles faithfully
describes the approach to the bifurcation, and may hence serve as a
starting point to further explore the nucleation of droplets and its
feedback on the transport of supersaturation.

\begin{figure}
\[ 
   \includegraphics{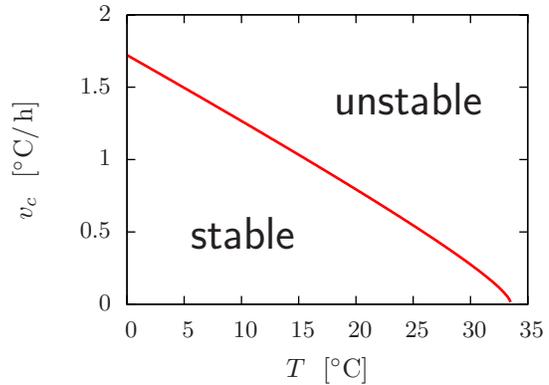} 
\]
\vspace*{-6mm}
\caption[max heating rate]{(Color online) 
  The solid red line gives an estimate~\eq{v_c} where the saddle node
  bifurcation leads to a breakdown of continuous diffusive
  demixing in the mixtures of hexane and methanol considered in
  detail in Ref.~\cite{auernhammer05JCP}. 
  The label `{\sffamily{stable}}' marks the range of heating rates $v$
  where diffusion keeps pace with the generation of supersaturation by
  the change of temperature. For larger heating rates (region marked
  as `{\sffamily{unstable}}') supersaturation will become so large that
  droplets are nucleated. 
  \label{fig:heatRate}}
\end{figure}

We conclude the present study by estimating the maximal cooling rates
where a mixture can still catch up with the change of temperature in a
system of lateral extension $\Lambda \simeq 1\;$cm. 
For a mixture of methanol and hexane the equilibrium diffusion
coefficient at the considered temperatures is of the order of $D
\simeq 10^{-5}$cm$^2$/s.  In view of \Eqs{xi} and \eq{defN} the system
can only smoothly follow the change of temperature when the cooling
rate the cooling rate does not exceed a critical value
\begin{equation}
v_c 
\equiv \left. \frac{\rmd T}{\rmd t} \right|_c
\simeq 
\frac{N_c}{ \6_T\phi_0 } \; \frac{D}{\Lambda^2}
\label{eq:v_c}
\end{equation}
By comparison with the phase diagram for the mixtures (\cf Fig.~5 in
\cite{auernhammer05JCP}) one finds that for all experimentally
accessible temperatures the cooling rate must not exceed a few K/h
(\Fig{heatRate}).
%
%
For other binary mixtures these numbers will be of the same order, and
for polymer solutions where the diffusion coefficient is still three
orders of magnitude smaller the cooling rate must not exceed a few
Kelvin per month. 

in view of the results summarized in \Fig{heatRate} this study
strongly suggest that complex time-dependent nonlinear behavior must
be considered as the typical response of a binary liquid which is
driven into the coexistence region by a temperature ramp. Only when
special precautions are taken, or when one approaches a region in the
phase diagram where temperature changes do not induce changes of the
composition diffusion can keep the system close to equilibrium.

\begin{acknowledgments}
  
  I am grateful to Doris Vollmer and G\"unter Auernhammer for many
  friutful discussions and providing me with the data on the
  methanol-hexane phase diagram needed to generate \Fig{heatRate}.
  In addition, I would like to thank Frank Dettenrieder, Bruno
  Eckhardt, Siegfried Grossmann, and Burkhard D\"unweg for very useful 
  discussions, and 
  Stephan Herminghaus and Martin Brinkmann for feedback on a draft of 
  this manuscript. 
%
\end{acknowledgments}

\appendix

\section{Asymmetric phase diagrams}
\label{app:asymm}

\begin{figure}
\[ \includegraphics{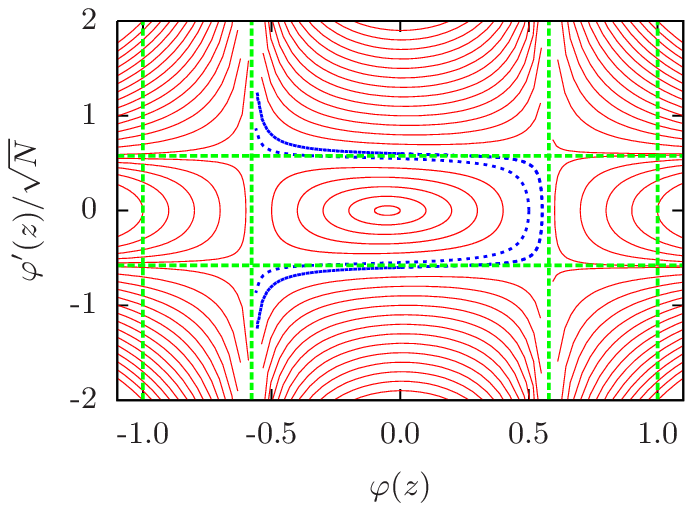} \]
\vspace*{-6mm}
\caption[phase flow]{(Color online) 
  Flow diagram of \Eq{profileAsym} for $\zeta/\xi = 0.067$.
  All other parameters are the same as in \Fig{phaseFlow}.
\label{fig:asymmPhaseFlow}}
\end{figure}

In \cite{auernhammer05JCP} it is shown that binary mixtures with a
symmetric coexistence region are described by \Eq{profile}, even
when the critical point does not correspond to equal volume fractions
of both phases. An asymmetric miscibility gap can be characterized by
the temperature dependence of the mean value $\bar\phi(T)$ of the
coexisting compositions. It gives rise to an additional source term
\(
\zeta \equiv \phi_0^{-1}\,\6_t\bar\phi
\)
in the diffusion equation, which breaks the $\vphi\to -\vphi$
symmetry. Absorbing $\xi\Lambda^2/D$ again into the length scale and
concentrating on the case $M\ll 1$ (\ie $\delta\ll\Lambda$) one obtains 
\begin{equation}
   \6_z\left[ \frac{3\vphi^2-1}{2} \: \6_z\vphi \right]
 \simeq
   \vphi - \frac{\zeta}{\xi}.
\label{eq:profileAsym}
\end{equation}
The flow diagram for the stationary concentration profiles for
$\zeta/\xi = 0.067$ is shown in \Fig{asymmPhaseFlow}. In comparison to
the symmetric case \Fig{phaseFlow}
the heteroclinic connection between the saddle points on the spinodal
disappear. This leads to a rearrangement of the flow in the unstable
region $-\vphi_s<\vphi<\vphi_s$.  On the other hand, there are only
minor changes in the profiles outside this range, and the criterion
$|\vphi'(0)|\leq 1$ to obtain a stable diffusive profile remains valid
to a very good accuracy.

\section{Mode expansion for the diffusion equation}
\label{app:modeExpansion}

When inserting the ansatz \eq{expansion} into the nonlinear diffusion
equation \eq{diffu} only the nonlinear terms take some effort to
evaluate. 
We observe that
\[
\6_z\left[ \frac{3\vphi^2-1}{2} \: \6_z\vphi \right]
=
  3 \vphi \: ( \6_z\vphi )^2 
+ \frac{3}{2} \vphi^2 \: \6_z^2\vphi 
- \frac{1}{2} \6_z^2\vphi 
\]
and evaluate the terms one after the other. 

Observing that 
\(
  \6_z\vphi
= \sum_j' j \, \vphi_j \, \cos \frac{\pi j z}{2}
\)
we thus obtain
\begin{widetext}
\begin{eqnarray*}
  3 \vphi \: ( \6_z\vphi )^2 
& = &
\frac{6}{\pi} \;
{\sum_{jkl}}' k l \; \vphi_j  \, \vphi_k  \, \vphi_l  \; 
\sin \frac{\pi j z}{2} \; \cos \frac{\pi k z}{2} \; \cos \frac{\pi l z}{2}
\\[2mm]
& = &
\frac{3}{2 \pi} \;
{\sum_{jkl}}' k l \; \vphi_j  \, \vphi_k  \, \vphi_l  \; 
\left[
  \sin \frac{\pi (j+k+l) z}{2} 
+ \sin \frac{\pi (j-k-l) z}{2} 
+ \sin \frac{\pi (j+k-l) z}{2} 
+ \sin \frac{\pi (j-k+l) z}{2} 
\right]
\end{eqnarray*}
\end{widetext}
Using the antisymmetry of $\vphi_j$ and shifting the summation index
this yields
\begin{subequations}
\begin{eqnarray}
  3 \vphi \: ( \6_z\vphi )^2 
& = &
\frac{6}{\pi} \;
{\sum_{jkl}}' k l \; \vphi_j  \, \vphi_k  \, \vphi_l  \; 
  \sin \frac{\pi (j+k+l) z}{2} 
\nonumber
\\[2mm]
& = &
\frac{6}{\pi} \;
{\sum_{jkl}}' k l \; \vphi_{j-k-l}  \, \vphi_k  \, \vphi_l  \; 
  \sin \frac{\pi j z}{2} 
\end{eqnarray}
By an analogous calculation one finds
\begin{equation}
\frac{3}{2} \vphi^2 \: \6_z^2\vphi 
=
\frac{6}{\pi} \;
{\sum_{jkl}}' \frac{k^2}{2} \; \vphi_{j-k-l}  \, \vphi_k  \, \vphi_l  \; 
  \sin \frac{\pi j z}{2} 
\end{equation}
\end{subequations}
Collecting the expansion coefficients of the Fourier series then
immediately leads to \Eq{dot_vphi}.



\end{document}